\documentclass[letter]{aa} 
\usepackage{graphicx}
\usepackage{txfonts}
\usepackage{natbib}
\bibpunct{(}{)}{;}{a}{}{,} 

\begin{document}
\title{Angular momentum transport in stellar interiors constrained by rotational splittings of mixed modes in red giants}
\titlerunning{Angular momentum transport and rotational splittings in red giants}

\author{P. Eggenberger\inst{1} \and J. Montalb\'{a}n\inst{2} \and A. Miglio\inst{3}}

   \offprints{P. Eggenberger}

\institute{Observatoire de Gen\`eve, Universit\'e de Gen\`eve, 51 Ch. des Maillettes, CH-1290 Sauverny, Suisse 
\and
  Institut d'Astrophysique et de G\'eophysique de l'Universit\'e de Li\`ege, 17 All\'ee du 6 Ao\^ut, B-4000 Li\`ege, Belgique
\and
School of Physics and Astronomy, University of Birmingham, Edgbaston, Birmingham B15 2TT
       }

   \date{Received; accepted}

 
  \abstract
   {Recent asteroseismic observations have led to the determination of rotational frequency splittings for $\ell=1$ mixed modes in red giants.}
   {We investigate how these observed splittings can constrain the modelling of the physical processes transporting angular momentum in stellar interiors.} 
   {We first compare models including a comprehensive treatment of shellular rotation only, with the rotational splittings observed for the red giant \object{KIC 8366239}. We then study how these asteroseismic constraints can give us information about the efficiency of an additional mechanism for the internal transport of angular momentum. This is done by computing rotating models of \object{KIC 8366239} that include a constant viscosity corresponding to this physical process, in addition to the treatment of shellular rotation.}
   {We find that models of red giant stars including shellular rotation only predict steep rotation profiles, which are incompatible with the measurements of rotational splittings in the red giant \object{KIC 8366239}. Meridional circulation and shear mixing alone are found to produce an insufficient internal coupling so that an additional mechanism for the internal transport of angular momentum is needed during the post-main sequence evolution. We show that the viscosity $\nu_{\rm add}$ corresponding to this mechanism is strongly constrained to be $\nu_{\rm add}=3 \times 10^{4}$\,cm$^2$\,s$^{-1}$ thanks to the observed ratio of the splittings for modes in the wings to those at the centre of the dipole forests. Such a value of viscosity may suggest that the same unknown physical process is at work during the main sequence and the post-main sequence evolution.}
   {}


   \maketitle
%

\section{Introduction}

Rotation is an important process that can have a significant impact on the evolution and properties of stars \citep[e.g.][]{mae09}. The effects of rotation and in particular the transport of chemicals and angular momentum by meridional circulation and shear instability have been included in stellar evolution codes \citep[see the review by][]{mae12}. This has generally been done in the context of shellular rotation, which assumes that a strongly anisotropic turbulence leads to an essentially constant angular velocity on the isobars \citep{zah92}. To progress in our understanding of the dynamical processes at work in stellar interiors, one needs to confront these theoretical models with observational constraints. 

Surface abundances are used as diagnostics of the efficiency of rotational mixing in stellar interiors. In the case of low-mass stars, measurements of light element abundances are particularly valuable for this purpose \cite[see e.g.][]{tal10, pin10}. Rotation is indeed found to have an impact on the lithium depletion of a solar-type star from the very beginning of its evolution \citep{egg12} to more advanced phases as a red giant \citep{pal06}. Interestingly, the predicted lithium abundance at the surface of main-sequence stars is very sensitive to the modelling of rotation. While lithium abundances observed for stars on the blue side of the lithium dip can be reproduced by rotating models including shellular rotation \citep{pal03}, the same models fail at reproducing lithium abundances for stars on the cool side of the dip \citep{tal98,tal03}. This indicates that there is an additional physical process operating in the internal layers of low-mass stars, which are efficiently spun down via magnetic braking during their evolution on the main sequence \citep{tal98,tal03,cha05}.

Additional constraints of the modelling of angular momentum transport in stellar interiors come from measurements of the surface rotation of stars. 
Observations of rotational periods for young solar-type stars in open clusters suggest that slow rotators develop a high degree of differential rotation between the radiative core and the convective envelope, while solid-body rotation is favoured for fast rotators \citep[][]{irw07,bou08,den10_spin,den10}. Models including a detailed treatment of shellular rotation show however that meridional circulation and shear mixing alone do not couple the core and envelope, even in the presence of rapid rotation \citep{egg10_magn}. This also indicates the need to introduce an additional mechanism for the transport of angular momentum during the beginning of the main-sequence evolution of solar-type stars.

In addition to observations of stellar surface properties, those of
the solar five-minute oscillations have provided a wealth of information
on the internal structure of the Sun and have led to the determination of its rotation profile \citep{bro89,kos97,cou03,gar07}. This major constraint on the modelling of angular momentum transport shows that rotating models including meridional circulation and shear turbulence are unable to reproduce the near uniformity of the solar rotation profile \citep{pin89,cha95,egg05_mag,tur10}. 
The helioseismic results stimulated various attempts to obtain similar observations for stars with different masses and at various evolutionary stages. In the past few years, the spectrographs developed for extra-solar planet searches have enabled the detection of solar-like oscillations for a handful of solar-type stars \citep[see e.g.][]{bed08}, as well as for a few red giants \citep{fra02, bar04, bar07, der06}. Owing to the limited duration of these ground-based asteroseismic campaigns, no clear additional constraints have however been obtained about the internal transport of angular momentum, with only one tentative evidence of rotational splittings being reported for the F9 V star $\beta$~Virginis \citep{car05_bvir,egg06}. We recall that rotation lifts the degeneracy in the azimuthal order $m$ of non-radial oscillation modes. This results in $(2\ell+1)$ frequency peaks in the power spectrum for each mode. The frequency spacings between these peaks are called rotational splittings and are directly related to the angular velocity and the properties of the modes in their propagation regions inside the star.

With the launch of the CoRoT \citep{bag06} and Kepler \citep{bor10} space missions, solar-like oscillations are now being detected and characterized for a very large number of stars. Photometric measurements of solar-like oscillations with the Kepler spacecraft have recently led to the precise determination of the rotational frequency splittings of mixed modes in red giants \citep{bec12}. Mixed modes are identified in the power spectrum as dense clusters of modes referred to as `dipole forests' for $\ell=1$ modes. The mode at the centre of such a forest is dominated by its acoustic character and is more sensitive to the external layers, while adjacent modes in the `wings' of the forest are more gravity-dominated and sensitive to the stellar core. In this work, we investigate which constraints can be brought by the observations of \cite{bec12} to the theoretical modelling of the physical processes transporting angular momentum in stellar interiors.

The comparison between theoretical rotational splittings obtained with models including shellular rotation only and the splittings observed for the red giant \object{KIC 8366239} \citep{bec12} is first discussed in Sect.~2. Constraints imposed by these observed rotational splittings on an additional mechanism for the transport of angular momentum in stellar interiors are studied in Sect.~3, while our conclusions are given in Sect.~4.

\section{Models including shellular rotation only}
\label{sec_rot}

We use the Geneva stellar evolution code that includes a detailed treatment of shellular rotation \citep{egg08}. The basic physical ingredients of numerical models of rotating stars are described in Sect.~2 of \cite{egg10_melange}. We simply recall here that the transport of angular momentum obeys the equation
\begin{equation}
  \rho \frac{{\rm d}}{{\rm d}t} \left( r^{2}\Omega \right)_{M_r} 
  =  \frac{1}{5r^{2}}\frac{\partial }{\partial r} \left(\rho r^{4}\Omega
  U(r)\right)
  + \frac{1}{r^{2}}\frac{\partial }{\partial r}\left(\rho D r^{4}
  \frac{\partial \Omega}{\partial r} \right) \, , 
\label{transmom}
\end{equation}
where $r$ is a characteristic radius, $\rho$ the mean density on an isobar, $\Omega(r)$ the mean angular velocity at level $r$, and $U(r)$ the vertical component of the meridional circulation velocity. The first term on the right-hand side of Eq. (\ref{transmom}) describes the advection of angular momentum by meridional circulation. In the classical context of shellular rotation, both meridional circulation and shear mixing are considered as the main mixing mechanisms in radiative zones. The diffusion coefficient $D$ then only accounts for the transport of angular momentum by shear instability with $D=D_{\rm shear}$. In convective zones, solid-body rotation is assumed to occur, in accordance with the solar case.

\begin{figure}[htb!]
\resizebox{\hsize}{!}{\includegraphics{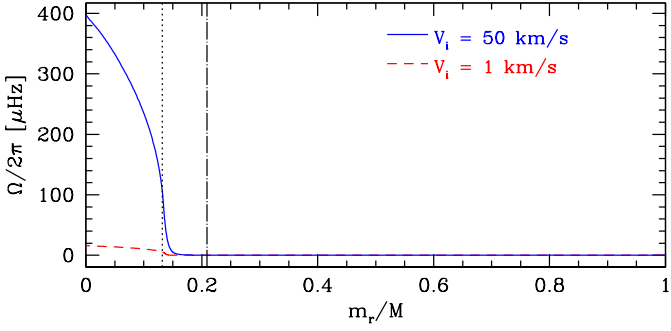}}
 \caption{Rotation profiles for models of \object{KIC 8366239} computed with shellular rotation only. The continuous and dashed lines correspond to an initial velocity on the ZAMS of 50\,km\,s$^{-1}$ and 1\,km\,s$^{-1}$. The vertical dotted and dashed-dotted lines indicate the locations of the border of the helium core and the convective envelope, respectively.}
  \label{pro_om_rot}
\end{figure}

A modelling of the red giant \object{KIC 8366239} is performed by computing rotating models including shellular rotation only. These models are evolved with the solar chemical composition given by \cite{gre93} and a solar calibrated value for the mixing-length parameter. We first adopt an initial velocity of 50\,km\,s$^{-1}$ on the zero-age main sequence (ZAMS) and obtain a model with an initial mass of 1.5\,M$_{\odot}$ in the H-shell burning phase that reproduces the global asteroseismic properties of \object{KIC 8366239} (i.e the large separation and the frequency of maximum oscillation power). This model is characterised by a mass of 1.4993\,M$_{\odot}$, a luminosity of 12.59\,L$_{\odot}$, a radius of 5.27\,R$_{\odot}$ and a mass of the inert helium core of 0.198\,M$_{\odot}$. The rotation profile of this model is shown in Fig.~\ref{pro_om_rot}. As expected from previous computations of rotating models including shellular rotation, the rotation profile during the red giant phase is very steep as a result of the central contraction occurring at the end of the main sequence \citep[][]{pal06,egg10_rg}.

The rotational splittings of the dipole modes ($\ell=1$) are then computed for this model using the Aarhus adiabatic 
pulsation code \citep{chr08}. A mean value of 32.8\,$\mu$Hz is then found for modes at the centre of the dipole forests. This is more than three orders of magnitude higher than the mean value of $0.135 \pm 0.008$\,$\mu$Hz observed in \object{KIC 8366239}. This huge discrepancy between observed and theoretical mean values of rotational splittings mainly comes from the large increase in the rotational velocity in the core of the model. Moreover, the ratio of the average rotational splitting of the first modes in the wings of the dipole forest to the mean splitting of the forest centre is equal to 2.5, which is significantly larger than the observed value of 1.5. We recall here that all mixed $\ell=1$ modes are sensitive to the properties of the stellar core, but that they contain a different amount of pressure and gravity-mode influence. The mode at the centre of a dipole forest is most efficiently trapped in the acoustic cavity and is thus more sensitive to the properties of the stellar envelope than the other modes, which exhibit larger gravity components towards the wings of the dipole forest. The ratio of the rotational splittings of modes in the wings to those for modes at the centre of a dipole forest therefore gives important constraints on the difference between the rotation rates of the stellar core and the outer layers.
The discrepancy between theoretical and observed values of this ratio indicates that models including shellular rotation only, predict rotation profiles during the red giant phase that are too steep to correctly reproduce the measurements of rotational splittings in \object{KIC 8366239}. We must however emphasize that the computation of rotational splittings is done here by using a standard first-order perturbation expression. While the assumption of slow rotation is valid in the envelope of our model, it is questionable in the very central layers, where the angular velocities become comparable to the frequencies of the oscillation modes. One can then wonder whether the large difference found between theoretical and observed rotational splittings is really related to the too-steep rotation profiles predicted by models including shellular rotation only, and is not simply due to an improper computation of these splittings for a model with a rapidly rotating core.

To investigate this point, a model of \object{KIC 8366239} is computed with a low initial velocity on the ZAMS of 1\,km\,s$^{-1}$. The corresponding rotation profile is shown in Fig.~\ref{pro_om_rot}. A decrease in the initial velocity leads to a global decrease in the internal rotation rates. This results in a strong decrease in the rotational splittings for modes at the centre of the dipole forests with a mean value of 1.6\,$\mu$Hz. This is still however more than one order of magnitude larger than the observed value. Moreover, the ratio of the splittings of modes in the wings to those for modes at the centre of dipole forests remains significantly larger than observed (2.4 instead of 1.5). The angular rotational velocities in the central parts of this model are much lower than the modes frequencies so that the expression used for the computation of rotational splittings is perfeclty valid in this case. A slow initial rotation is thus found to produce rotation profiles during the red giant phase that are also too steep compared to observations. We conclude that meridional circulation and shear instability alone produce an insufficient coupling to account for the rotational splittings observed in the red giant \object{KIC 8366239}, and that an additional mechanism for the transport of angular momentun must operate in stellar interiors during the post-main sequence evolution.

 \begin{figure}[htb!]
 \resizebox{\hsize}{!}{\includegraphics{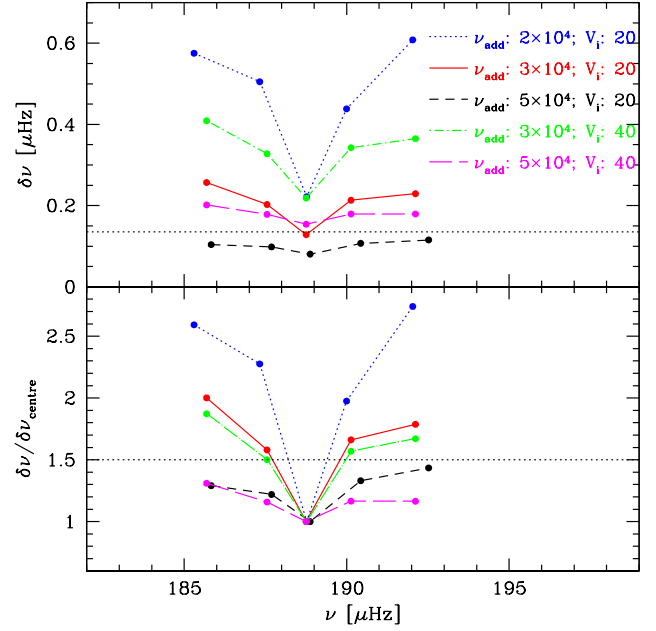}}
  \caption{Rotational splittings $\delta \nu$ for modes of the dipole forest near 190\,$\mu$Hz for models of \object{KIC 8366239} computed with different initial velocities on the ZAMS ($V_{\rm i}$ in km\,s$^{-1}$) and viscosities ($\nu_{\rm add}$ in cm$^2$\,s$^{-1}$) corresponding to an additional mechanism for the transport of angular momentum in radiative zones. The bottom panel shows the rotational splittings normalised to the splitting of the mode at the centre of the dipole forest. The dotted horizontal lines indicate the observed values of both the mean rotational splittings of modes at the centre of the dipole forests (top pannel) and the ratio of the splittings of the first modes in the wings of the dipole forest to the splitting of the mode at the centre of the forest (bottom panel).}
  \label{split_nu}
\end{figure}

\begin{figure}[htb!]
 \resizebox{\hsize}{!}{\includegraphics{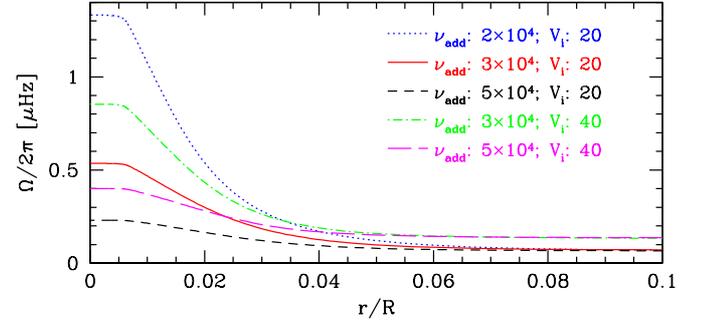}}
  \caption{Rotation profiles in the central layers for the models shown in Fig.~\ref{split_nu}.}
  \label{pro_om_nuadd}
\end{figure}

\section{Constraints on an additional mechanism for internal angular momentum transport}

We now investigate in more detail how the rotational splittings observed in \object{KIC 8366239} can constrain the needed additional mechanism for the transport of angular momentum. For this purpose, we introduce a new viscosity $\nu_{\rm add}$ corresponding to an additional physical process for the transport of angular momentum in radiative interiors. The total diffusion coefficient in Eq. (\ref{transmom}) is then given by $D=D_{\rm shear}+\nu_{\rm add}$.
The additional viscosity is not introduced into the equation describing the transport of chemical elements. This corresponds well to the two main mechanisms currently proposed to efficiently extract angular momentum from the central core of a solar-type star -- magnetic fields and internal gravity waves -- having a strong impact on the transport of angular momentum and only an indirect effect on the transport of chemical elements through the change of the rotation profile.
As a first step to progress in our understanding of this additional physical process, we simply assume that $\nu_{\rm add}$ is constant. We then investigate how its value can be constrained by asteroseismic observations of rotational splittings in the red giant \object{KIC 8366239} and thus determine the efficiency of this mechanism.

A small grid of rotating models including this additional mechanism are then computed for different initial velocities on the ZAMS ($V_{\rm i}$) and values of the viscosity ($\nu_{\rm add}$). To correctly reproduce the rotational splittings observed in \object{KIC 8366239}, a viscosity $\nu_{\rm add}=3 \times 10^{4}$\,cm$^2$\,s$^{-1}$ is found with an initial velocity on of 20\,km\,s$^{-1}$. Figure~\ref{split_nu} shows the comparison between the theoretical rotational splittings for this model and the observed values obtained for \object{KIC 8366239}, while the corresponding rotation profile is shown in Fig.~\ref{pro_om_nuadd}. We note that only the splittings near 190\,$\mu$Hz are shown in Fig.~\ref{split_nu}, but that the results are identical for the splittings observed near 160, 175, and 200\,$\mu$Hz.

We find that the two input parameters of our models ($V_{\rm i}$ and $\nu_{\rm add}$) can be precisely determined thanks to the simultaneous constraints coming from the observed values of the mean rotational splitting and from the ratio of splittings for modes in the wings to those for modes at the centre of the dipole forest. The bottom panel of Fig.~\ref{split_nu} shows that the aforementioned ratio enables a direct and precise determination of $\nu_{\rm add}$. As shown in Fig.~\ref{pro_om_nuadd}, an increase in $\nu_{\rm add}$ results in a more efficient transport of angular momentum, hence a flatter rotation profile in the radiative zone. Since the ratio of splittings for modes in the wings to those at the centre of the dipole forest is sensitive to the difference between the rotation rates of the stellar core and the outer layers, it is found to significantly decrease when the viscosity increases. The observed values of 1.5 for the first modes in the dipole forest is thus found to strongly constrain $\nu_{\rm add}$ to $3 \times 10^{4}$\,cm$^2$\,s$^{-1}$, independently of the initial velocity adopted in our models. The initial velocity $V_{\rm i}$ of the model can then be obtained thanks to the value of the rotational splitting of the modes at the centre of the dipole forest. For a given viscosity, this value increases with the initial velocity as a result of the global increase in the rotation rates within the stellar interior. As shown in the top panel of Fig.~\ref{split_nu}, the observed rotational splitting of $0.135 \pm 0.008$\,$\mu$Hz for modes at the centre of a dipole forest gives an initial velocity on the ZAMS of 20\,km\,s$^{-1}$ (for the viscosity of $3 \times 10^{4}$\,cm$^2$\,s$^{-1}$, which is determined from the ratio of splittings for modes in the wings to those for modes at the centre of the dipole forest). The value of 20\,km\,s$^{-1}$ found for the initial velocity on the ZAMS depends of course on the assumption made about the surface braking during the main sequence. A stronger braking results in a larger loss of angular momentum during the main sequence so that a higher value of the initial velocity is then required to correctly reproduce the observed mean value of rotational splittings. We note that the value obtained for the viscosity is insensitive to the adopted braking, because it is directly contrained by the ratio of splittings for modes in the wings to those for modes at the centre of the dipole forest, independently of the mean value of the splittings. We however recall here that the value of $3 \times 10^{4}$\,cm$^2$\,s$^{-1}$ found for the viscosity in the radiative interior has been obtained by using the usual assumption of solid-body rotation in convective zones.

We conclude that the efficiency of the additional mechanism for the transport of angular momentum in radiative zones operating during the post-main sequence evolution, can be strongly constrained by the observation of rotational splittings for modes in the wings of the dipole forests for the red giant \object{KIC 8366239}. The comparison between the determined viscosity of $3 \times 10^{4}$\,cm$^2$\,s$^{-1}$ and the coefficient $D_{\rm shear}$ shows that similar values are found for $D_{\rm shear}$ at the border of the convective envelope, but that this coefficient rapidly decreases towards the centre of the star with typical values of lower than $50$\,cm$^2$\,s$^{-1}$ in the stellar core. This underlines the need to introduce a new transport mechanism to efficiently transport angular momentum in the central parts of the star to obtain a core rotating at a moderate rate during the red giant phase.
Interestingly, the viscosity of $3 \times 10^{4}$\,cm$^2$\,s$^{-1}$ deduced from asteroseismic observations of a red giant is similar to the value 
required to correctly account for the observed spin-down of slow rotating solar-type stars during the beginning of the main sequence \citep{den10_spin}. This suggests that the same unknown physical process for the internal transport of angular momentum may be operating during the main-sequence and the post-main sequence evolution.

\section{Conclusion}

We have found that models of red giant stars including shellular rotation only predict steep rotation profiles in the central stellar layers, which are incompatible with observational measurements of rotational splittings in the red giant \object{KIC 8366239}. Asteroseismic observations indeed show that meridional circulation and shear instability alone produce insufficient internal coupling and that an additional mechanism for the internal transport of angular momentum is thus needed during the post-main sequence evolution.

We then investigated in more detail how the rotational splittings detected in \object{KIC 8366239} can be used to characterize this physical process by computing stellar models including an additional viscosity related to this mechanism. We found that the ratio of splittings for modes in the wings to those for modes at the centre of the dipole forests strongly constrains the efficiency of this additional physical process and leads to a precise determination of the corresponding viscosity of $3 \times 10^{4}$\,cm$^2$\,s$^{-1}$. Such a value of viscosity may indicate that the same unknown physical process for internal angular momentum transport is at work during the main-sequence and the red giant phase.

\begin{acknowledgements}
This work was partly supported by the Swiss National Science Foundation.
\end{acknowledgements}

\bibliographystyle{aa} 
\bibliography{biblio} 

\end{document}